\documentclass[aps,prl,reprint,showpacs]{revtex4-1}
\usepackage{amsmath,amsfonts,amssymb,bbm,graphicx,color,epstopdf}
\newcommand{\1}{\mathbbm 1}

\begin{document}

\title{Conserving and Gapless Hartree-Fock-Bogoliubov theory for 3D dilute
Bose gas}
\author{Ya-Hui Zhang and Dingping Li}
\email{lidp@pku.edu.cn}
\affiliation{Department of Physics, Peking University, Beijing, 100871,
China}
\begin{abstract}
The excitation spectrum for the three dimensional Bose gas in Bose-Einstein
Condensation phase is calculated nonperturbatively with Modified
Hartree-Fock-Bogoliubov theory, which is both conserving and gapless. From
Improved $\Phi -$derivable theory, the diagrams needed to preserve
Ward-Takahashi Identity are resummed in a systematic and nonperturbative
way.  It is valid up to the critical temperature where the dispersion relation of the low
energy excitation spectrum changes from linear to quadratic. Because including the higher order fluctuation, the results show
significant improvement on the calculation of the shift of critical temperature with other conserving and gapless theories.
\end{abstract}

\maketitle












Since Einstein and Bose's first proposal of Bose-Einstein Condensation(BEC)
and experimental realization of it with alkali atoms\cite%
{Anderson14071995,Bradley1995-1687,Davis1995-3969}, weakly interacting
dilute Bose gas has attracted significant attentions\cite{Andersen}.

The description of BEC at zero temperature began from Bogoliubov\cite%
{Bogoliubov1947-4} and quantum loop corrections to energy density were
calculated up to two loops \cite%
{lee1957eigenvalues,wu1959ground,Hugenholtz1959,Braaten,LeeYang1958-,Glassgold}
at low temperature. The self consistent Hartree-Fock-Bogoliubov (HFB)
approximation was used but it gave a gapped spectrum\cite%
{Griffin1996,shi,hutchinson2000gapless}, violating the Hugenholtz-Pines
theorem\cite{Hugenholtz1959} or the Goldstone theorem\cite{Goldstone1962-},
which results from the spontaneous symmetry breaking of U(1). Popov theory
neglects the anomalous average and get a gapless spectrum\cite%
{popov2001functional}. But the anomalous average is not negligible at the
Broken Phase.  To correctly describe
the BEC at high temperature, we need a theory to be both
conserving(consistent with the conservation laws) and gapless\cite%
{hohenberg1965microscopic,Griffin1996,hutchinson2000gapless}.  The
many-body T -matrix has been used to obtain an modified Popov approximation%
\cite{bijlsma1997variational}. However, this approach yields the same
critical temperature as that of idea gas and the Hugenholz-Pines theorem is
not always satisfied(as noted in \cite{Andersen}). An improved Popov
approximation based on many-body T-matrix approximation was developed\cite%
{al2002low}\cite{MPopov} but its main application is in low dimensional
systems. Conserving and gapless approximation  has been developed by  T.Kita with modified Luttinger-Ward
functional
\cite{Kita2006,*Kita2009,*Kita2012}  and F.Cooper et.al. with leading-order auxiliary field approximation\cite{Cooper2010,*Cooper2011}. However, their results are still mean field like with infinite quasiparticle lifetime.

It is a challenge to develop a conserving and gapless theory beyond mean field level. It has been shown that the critical temperature($T_{c}$) of
weakly interacting Bose gas in 3 dimension is positively shifted from that
of idea gas ($T_{0}$) proportional to the scattering length a:$\frac{%
T_{c}-T_{0}}{T_{0}}=cn^{1/3}a$\cite{baym1999transition,*baym2000,*baym2001bose}. The accurate determination of $c$  by lattice simulations\cite{arnold2001bec,*kashurnikov2001critical} and other analytical calculations from uncondensed phase\cite{Kastening,*Ledowski,*de,*Kneur} shows $c\approx1.29$, while Kita and Cooper's theory from broken phase gives $c=2.33$\cite{Cooper2010,Kita2006}.
A non-perturbative theory beyond mean field level is needed to
correctly describe the broken phase near $T_{c}$. Besides, when interaction is
strong, the exact result may differ from mean field theory even at
low temperature due to strong fluctuation effects.

In this work, we presents a modified Hartree-Fock-Bogoliubov (MHFB)
approximation which is conserving and gapless and is beyond mean field level. We start from the gap equation of full HFB approximation by
removing the divergence due to the double counting. Then the infinite
series of diagrams needed to preserve the Ward-Takahashi Identity (WTI) are resummed. The approach in this paper is based on two particle irreducible(2PI) $\Phi$-derivable theory\cite{Luttinger,*de1964stationary,*Baym1961,*vanHees} and it
can be equivalently obtained from Schwinger-Dyson Equation approach\cite{rosenstein1989covariant,*QL,*okopinska1996goldstone}. The solution of
the Broken Phase ends at the temperature where the dispersion of low energy
excitation spectrum changes from linear to quadratic, which indicates a
second order phase transition. The critical temperature shift coefficient is $c=1.59$ and has significant improvement over Kita and Cooper's result. And because the method incorporates the
resummation of an infinite series of diagrams, the result differs from Popov
and Kita and Cooper theory and it can describe the damping of quasiparticles.

For Bose gas, the grand-canonical partition function can be written with
imaginary time path integral\cite{Andersen}:
\begin{equation}
Z[J,B]=\int D[\psi ,\psi ^{\ast }]e^{-S[\psi ^{\ast },\psi ]-\int
d(1)J_{i}\psi _{i}-\frac{1}{2}\int d(12)B_{ij}\psi _{i}\psi _{j}}
\label{partition}
\end{equation}%
Where the action (in dimensionless unit) is
\begin{equation}
S[\psi ,\psi ^{\ast }]=\int_{0}^{\beta }d\tau \int d^{d}x(\psi ^{\ast
}[\partial _{\tau }-\mu +\nabla ^{2}]\psi +\frac{g}{2}\psi ^{\ast }\psi
^{\ast }\psi \psi )  \label{action}
\end{equation}%
$g=8\pi a$, where a is the scattering length. $\beta =\frac{1}{T}$. $\psi
_{1}$, $\psi _{2}$ represent $\psi ^{\ast }$, $\psi $. $d(1)$ means $d\tau
d^{d}x$ and $J_{i}$, $B_{ij}$ are auxiliary sources which will be set zero
at last.

The 2PI (two particle irreducible) functional $\Gamma \lbrack \varphi ,G]$
is defined by the double Legendre transformation and can be written in the
form:
\begin{equation}
\Gamma \lbrack \varphi ,G]=S[\varphi _{i}]+\frac{1}{2}Tr\{D^{-1}(G-D)\}+%
\frac{1}{2}Tr\ln G^{-1}+\Phi \lbrack \varphi ,G]  \label{eq:thermo}
\end{equation}
where $D_{ij}^{-1}=\frac{\delta ^{2}S[\varphi _{i}]}{\delta \varphi
_{i}\varphi _{j}}$ and $\varphi _{i}=\left<\psi _{i}\right>$. $G_{ij}$ is
the Green Function $G_{ij}=\left\langle \psi _{i}\psi _{j}\right\rangle
_{c}=\left\langle \psi _{i}\psi _{j}\right\rangle -\left\langle \psi
_{i}\right\rangle \left\langle \psi _{j}\right\rangle $. $\Phi \lbrack
\varphi ,G]$ is the sum of all 2PI vacuum diagrams.

$\Phi \lbrack \varphi ,G]$ can be expanded to n loop and we get n loop $\Phi
$-derivable approximation. Then we can get the truncated $\varphi $ and $%
G^{tr}$ by solve:
\begin{equation}
\frac{\delta \Gamma \lbrack \varphi ,G^{tr}]}{\delta \varphi _{i}}=0\text{ ,
}\frac{\delta \Gamma \lbrack \varphi ,G^{tr}]}{\delta G_{ij}^{tr}}=0
\label{eq:gapeqa}
\end{equation}
Including the simplest diagrams (Hartree Fock approximation),%
\begin{equation}
\Phi \lbrack \varphi ,G]=\frac{g}{2}\int
d(1)[G_{11}(x,x)G_{22}(x,x)+2G_{12}(x,x)G_{21}(x,x)]
\end{equation}

For Homogeneous gas, we define $\upsilon =\varphi _{1}=\varphi _{2}$. Then $%
G_{ij}(x,y)=G_{ij}(x-y)$. $x$ means $(\tau ,\vec{x})$.

Then from (\ref{eq:gapeqa}) we get the shift equation and gap equation:
\begin{equation}
\mu =g\upsilon ^{2}+gG_{11}^{tr}(0)+2gG_{12}^{tr}(0)
\end{equation}
\begin{align}
\Gamma _{tr}^{(2)}& =  \notag \\
&
\begin{pmatrix}
\mathit{\ }\Sigma _{11}^{tr} & -i\omega _{n}-\mu +\Sigma _{12}^{tr}+k^{2} \\
i\omega _{n}-\mu +\Sigma _{12}^{tr}+k^{2} & \mathit{\ }\Sigma _{11}^{tr}%
\end{pmatrix}
\label{eq:reversegreen}
\end{align}

The equation is written after Fourier transformation $G_{ij}(x-y)=\frac{1}{%
V\beta }\sum_{\omega _{n,}k}G_{ij}(\omega _{n},\vec{k})e^{-i\omega _{n}\tau
+i\vec{k}\cdot (\vec{x}-\vec{y})}$; $\Gamma _{tr}^{(2)}=G^{tr-1}$. $\omega
_{n}$ is the Matsubara frequency $\omega _{n}=\frac{2\pi n}{\beta }$. And $%
\Sigma _{11}^{tr}=g\upsilon ^{2}+gG_{11}^{tr}(0)$, $\Sigma
_{12}^{tr}=2g\upsilon ^{2}+2gG_{12}^{tr}(0)$. Due to the symmetry, we have $%
G_{12}^{tr}(0)=G_{21}^{tr}(0),G_{11}^{tr}(0)=G_{22}^{tr}(0)$.

We define $\mu _{R}=\mu -\Sigma _{12}^{tr}$. $\mu _{R},\Sigma _{11}^{tr}$
can be solved self-consistently with $G^{tr}$ which is the inverse of $%
\Gamma ^{tr}$.
\begin{eqnarray}
\Sigma _{11}^{tr} &=&g\upsilon ^{2}+\frac{g}{V\beta }\sum_{\omega _{n,}k}%
\frac{\Sigma _{11}^{tr}}{(i\omega _{n})^{2}-\omega _{k}^{2}}  \notag \\
\mu _{R} &=&-2g\upsilon ^{2}+\Sigma _{11}^{tr}  \label{eq:gap}
\end{eqnarray}%
where, $\omega_{k}=\sqrt{(k^{2}-\mu _{R})^{2}-(\Sigma _{11}^{tr})^{2}}$.

In three dimension, $\alpha _{0}=\frac{1}{V\beta }\sum_{\omega _{n,}k}\frac{1%
}{(i\omega _{n})^{2}-\omega _{k}^{2}}=-\frac{1}{V}\sum_{k}(\frac{1}{\omega
_{k}}\frac{1}{e^{\beta \omega _{k}}-1}+\frac{1}{2\omega _{k}})$ has the
ultraviolet divergence due to the double counting problem, which arises
because we use the pseudopotential. The pseudopotential has already
effectively incorporated in the first term of the Born series the
information of the higher-order terms\cite{stoof2009ultracold}. To avoid
this problem, the vacuum terms should be subtracted: $\alpha _{R}=\alpha
_{0}+\frac{1}{V}\sum_{k}\frac{1}{2k^{2}}$. The equation (\ref{eq:gap}) after
renormalization is:
\begin{equation}
\Sigma _{11}^{tr}=g\upsilon ^{2}-g\frac{\Sigma _{11}^{tr}}{V}\sum_{k}(\frac{1%
}{\omega _{k}}\frac{1}{e^{\beta \omega _{k}}-1}+\frac{1}{2\omega _{k}}-\frac{%
1}{2k^{2}})  \label{eq:G11}
\end{equation}
The density $n=-\frac{1}{V}\frac{\partial \Omega }{\partial \mu }$ can be
calculated from (\ref{eq:thermo}):
\begin{equation}
n=\upsilon ^{2}+G_{12}^{tr}(0)  \label{eq:num}
\end{equation}
and
\begin{equation}
G_{12}^{tr}(0)=\frac{1}{V}\sum_{k}(\frac{k^{2}+\Sigma _{12}^{tr}}{\omega _{k}%
}\frac{1}{e^{\beta \omega _{k}}-1}+\frac{k^{2}-\mu ^{\prime }-\omega _{k}}{%
2\omega _{k}})  \label{eq:G12}
\end{equation}

We can get $\upsilon $ and $G^{tr}$ from $n$, $a$ and T with equations (\ref%
{eq:gap})(\ref{eq:G11})(\ref{eq:num})(\ref{eq:G12}).

WTI derived from 1PI formalism may be not preserved by $\Phi $-derivable
approximations due to the missing of some diagrams. An improved $\Phi $%
-derivable theory was developed to systematically add the missed diagrams%
. We use $%
\Gamma ^{{}}[\varphi ,G^{tr}]$ to approximate the 1PI effective action:
\begin{equation}
\Gamma ^{{}}[\varphi ]=\Gamma \lbrack \varphi ,G^{tr}(\varphi )]
\end{equation}
with $G^{tr}(\varphi )$ defined by $\frac{\delta \Gamma \lbrack \varphi
,G^{tr}]}{\delta G_{ij}^{tr}}=0$.

Because $\Gamma \lbrack \varphi ,G^{tr}(\varphi )]$ conserves the
symmetry(as in (\ref{eq:thermo})), the IPI effective action remains
unchanged under the transformation of U(1) symmetry. The Green Function
defined by the inverse of
\begin{equation}
\Gamma ^{(2)}=\frac{\Gamma ^{tr}[\varphi ]}{\delta \varphi _{i}\delta
\varphi _{j}}
\end{equation}
will be gapless. It's easy to show that
\begin{equation}
\Gamma ^{(2)}=\Gamma _{tr}^{(2)}+\frac{\delta ^{2}\Gamma \lbrack \varphi
,G^{tr}]}{\delta \varphi _{i}(x)\delta G_{mn}^{tr}}\frac{\delta G_{mn}^{tr}}{%
\delta \varphi _{j}(y)}
\end{equation}
$\frac{\delta G_{mn}^{tr}}{\delta \varphi _{j}(y)}$ can be got by taking the
derivative of:
\begin{equation}
\int d(2^{\prime })\Gamma _{tr;ij^{\prime }}^{(2)}G_{j^{\prime
}j}^{tr}=\delta _{ij}
\end{equation}
By defining $\Lambda _{j^{\prime }jm}^{tr}=\frac{\delta G_{j^{\prime }j}^{tr}%
}{\delta \varphi _{m}},\Gamma _{ijm}^{(3)}=\frac{\delta \Gamma _{tr;ij}^{(2)}%
}{\delta \varphi _{m}}$; we get
\begin{equation}
\Lambda _{j^{\prime }jm}^{tr}=-\int d(1^{\prime },2^{\prime })\Gamma
_{im^{\prime }m}^{(3)}G_{j^{\prime }i}G_{m^{\prime }j}  \label{BS}
\end{equation}
$\Gamma _{ijm}^{(3)}$ can be got by taking derivative of (\ref%
{eq:reversegreen}). These equations are actually the Bethe-Salpeter Equation
to solve $\Lambda _{j^{\prime }jm}^{tr}$.

In the level of HFB, we can get Modified HFB approximation:
\begin{eqnarray}
\Gamma _{11}^{(2)}(k) &=&\Sigma _{11}^{tr}+g\upsilon \Lambda
_{221}^{tr}(k)+2g\upsilon \Lambda _{121}^{tr}(k)  \notag \\
\Gamma _{12}^{(2)}(k) &=&-i\omega _{n}-\mu _{R}+k^{2}+g\upsilon \Lambda
_{222}^{tr}(k)+2g\upsilon \Lambda _{122}^{tr}(k)  \notag \\
\Gamma _{21}^{(2)}(k) &=&i\omega _{n}-\mu _{R}+k^{2}+g\upsilon \Lambda
_{111}^{tr}(k)+2g\upsilon \Lambda _{121}^{tr}(k)  \notag \\
\Gamma _{22}^{(2)}(k) &=&\Sigma _{11}^{tr}+g\upsilon \Lambda
_{112}^{tr}(k)+2g\upsilon \Lambda _{122}^{tr}(k)
\end{eqnarray}
Where $\Lambda _{mnl}^{tr}(k)$ is the Fourier transformation of $\Lambda
_{mnl}^{tr}(x,x,y)=\frac{\delta G_{mn}^{tr}(x,x)}{\delta \varphi _{l}(y)}$.
The latter can be solved by the Bethe-Salpeter Equation(\ref{BS}):
\begin{eqnarray}
\Lambda _{mnl}^{tr}(k) &=&\Lambda _{lll}^{tr}(k)I_{m\bar{l},\bar{l}%
n}(k)+\Lambda _{\bar{l}\bar{l}l}^{tr}(k)I_{ml,ln}(k)  \notag \\
&&+\Lambda _{\bar{l}ll}^{tr}(k)\left( 2I_{ml,\bar{l}n}(k)+2I_{m\bar{l}%
,ln}(k)\right)  \notag \\
&&+2\upsilon \left( I_{ml,\bar{l}n}(k)+I_{m\bar{l},ln}(k)+I_{m\bar{l},\bar{l}%
n}(k)\right)  \label{eq:self-con-k}
\end{eqnarray}%
where $\bar{l}$ is defined as $\delta _{l\bar{l}}=0$ and:
\begin{equation}
I_{mn,m^{\prime }n^{\prime }}(k)=-\frac{1}{V\beta }\sum_{\omega
_{n1},k_{1}}G_{mn}^{tr}(k_{1}+k)G_{m^{\prime }n^{\prime }}^{tr}(k_{1})
\label{eq:fish}
\end{equation}

Again due to the double counting, $I_{12,21}$ and $I_{21,12}$ have
ultraviolet divergences. They should be renormalized by subtracting vacuum
diagrams:
\begin{gather}
I_{12,21}^{R}=-[\frac{1}{V\beta }\sum_{\omega
_{n1},k_{1}}G_{12}^{tr}(k_{1}+k)G_{21}^{tr}(k_{1})-Vac]  \notag \\
Vac=\frac{1}{V}\sum_{k}\frac{1}{2k^{2}}
\end{gather}%
\begin{figure}[tbp]
\centering
\includegraphics[width=0.9\linewidth]{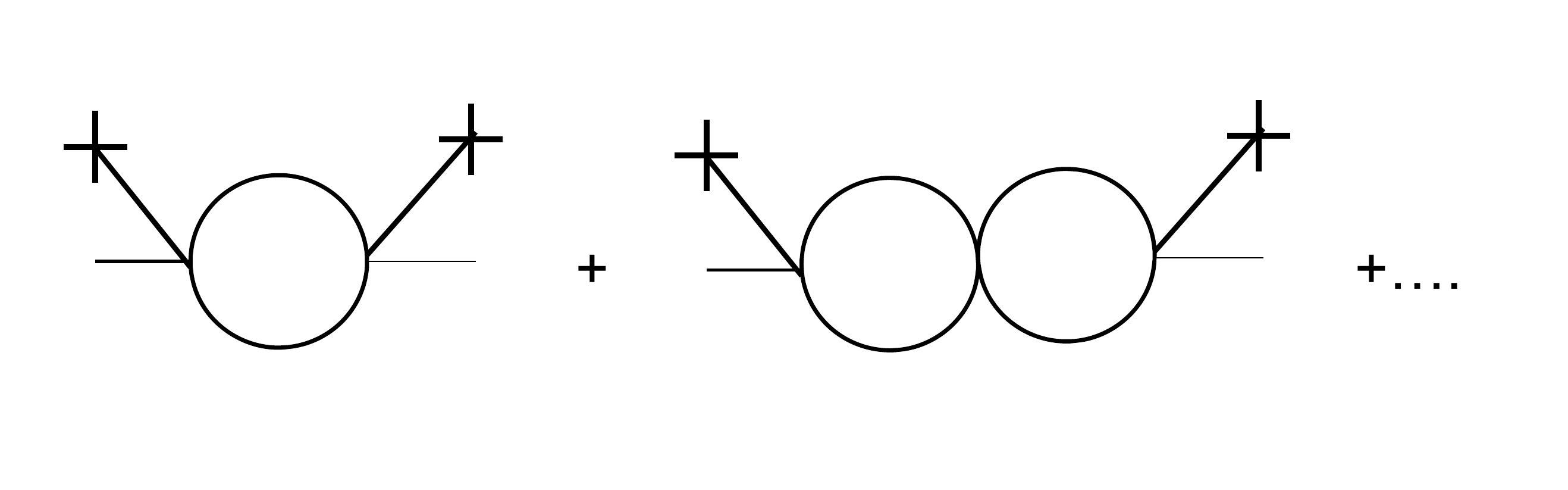}
\caption{The corrections to the self energy is the resummation of the
infinite series of diagrams. The propagator in the diagrams are the gapped HFB ones but the final propagator after resummation is gapless.}
\label{fig:bubble}
\end{figure}
Solve these linear equations and we can get the corrections to the self
energy. The corrections are the resummation of the infinite series of
diagrams (as shown in FIG.\ref{fig:bubble}.).

The Green Function is the inverse of $\Gamma ^{(2)}$. By analytic
continuation $i\omega _{n}\rightarrow\Omega +i\varepsilon $, the retarded
Green function $G^{R}$ is got and the spectral weight function is:
\begin{equation}
\rho (k,\Omega )=-2ImG^{R}(k,\Omega)
\end{equation}

We solve the gap equation numerically and the result is shown in FIG.\ref%
{fig:solution}. The equation ceases to have a solution at $T_{c}$, which is
the end point of the Broken Phase and is actually the critical point of a
second order phase transition. $\upsilon^2$ is not exactly equal to the condensation number
$n_0$ and needs corrections to get the exact $n_0$ just like that $G^{tr}$ needs corrections to get the exact 
Green function.

\begin{figure}[tbp]
\centering
\includegraphics[width=0.79\linewidth]{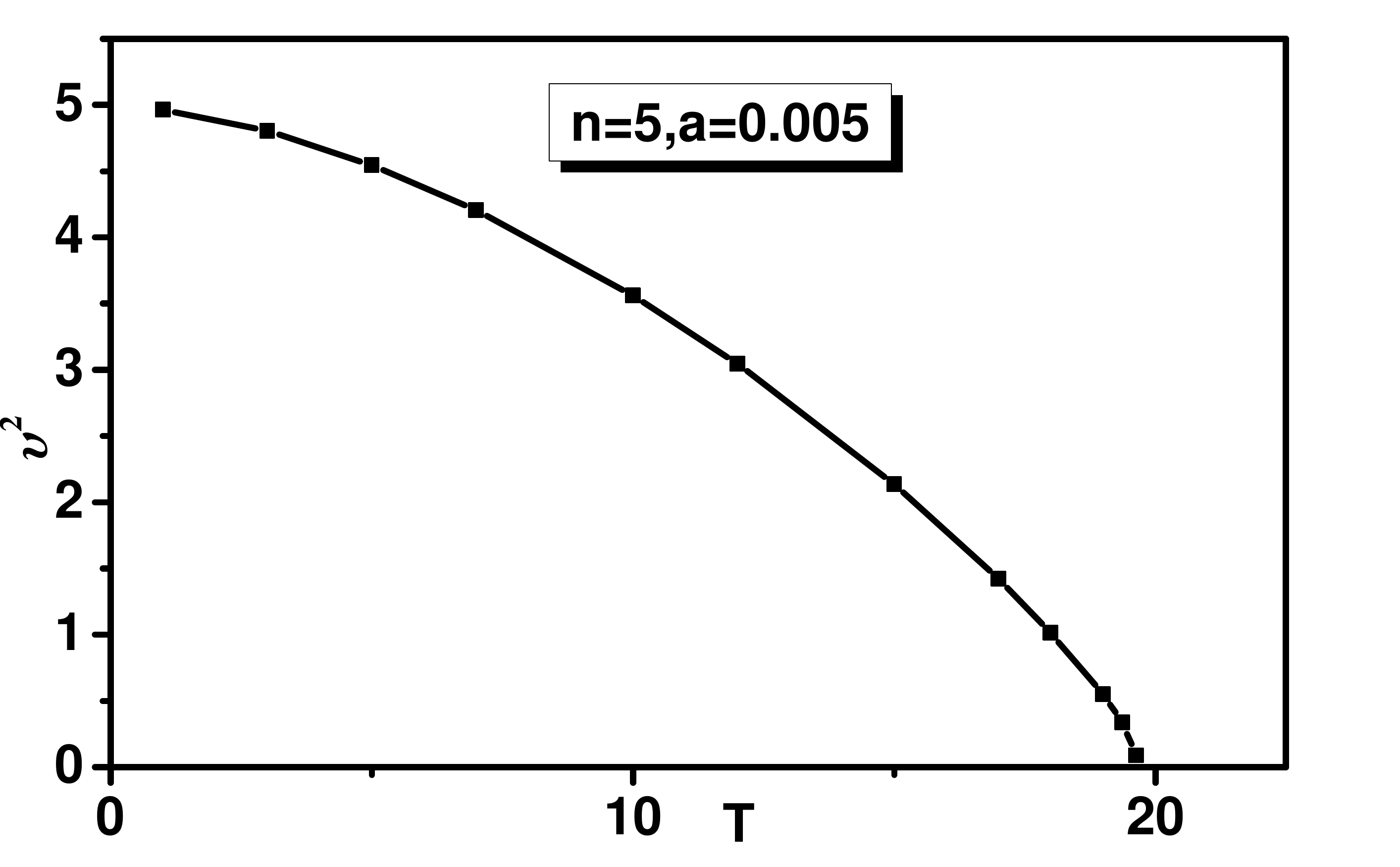}
\caption{$\upsilon^2-T$ for HFB theory when$ n=5,a=0.005$%
, $T_{0}=19.3716$, $T_{c}=19.635$. At $T_c$, $\upsilon^2=0.086$. }
\label{fig:solution}
\end{figure}

\begin{figure}[tbp]
\centering
\includegraphics[width=0.95\linewidth]{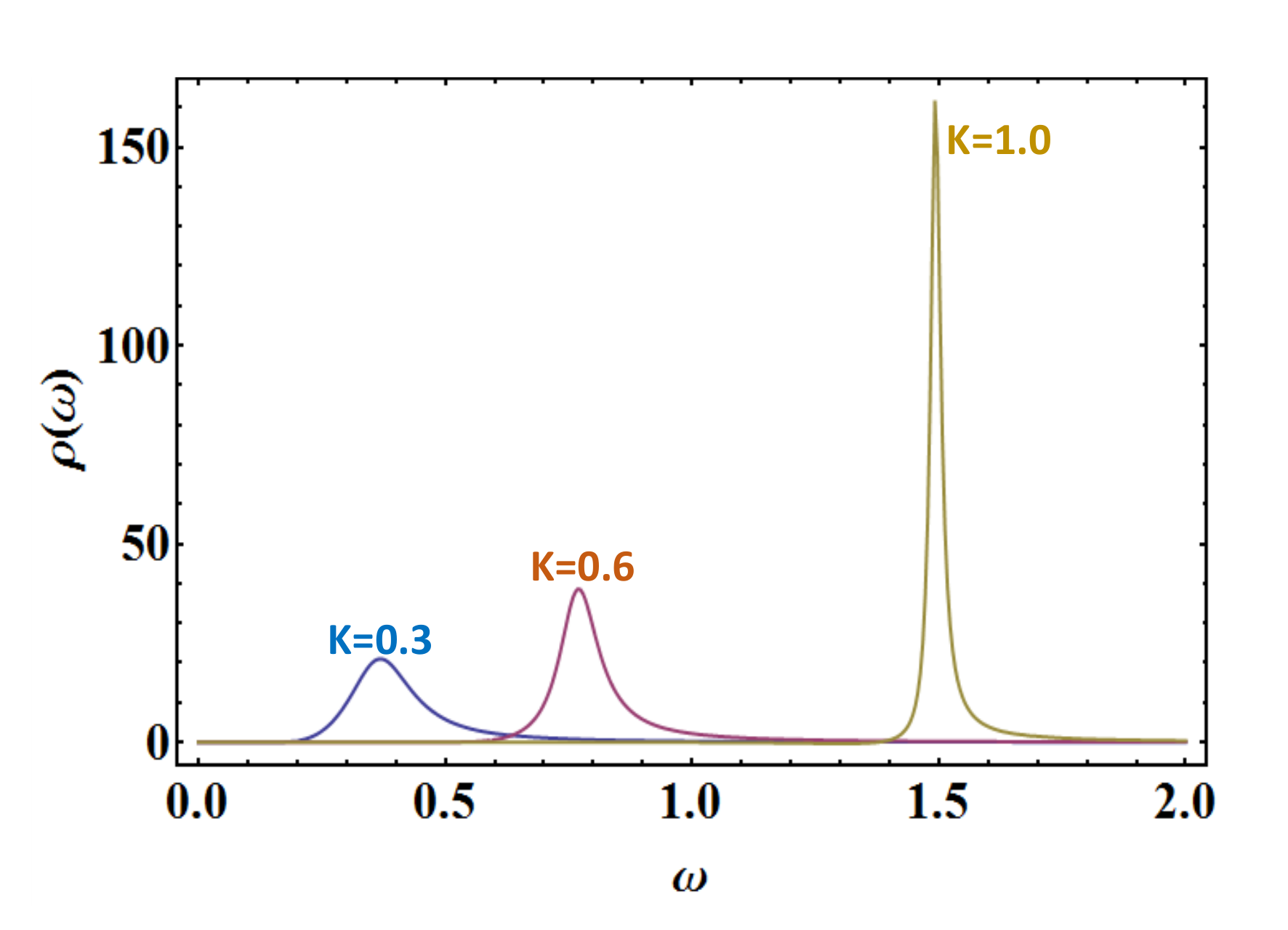}
\caption{The spectral weight for k=0.3, 0.6, 1 when n=5, a=0.005, T=10; It
is clear there is damping of quasiparticle.}
\label{fig:sw}
\end{figure}

\begin{figure}[tbp]
\centering
\includegraphics[width=0.95\linewidth]{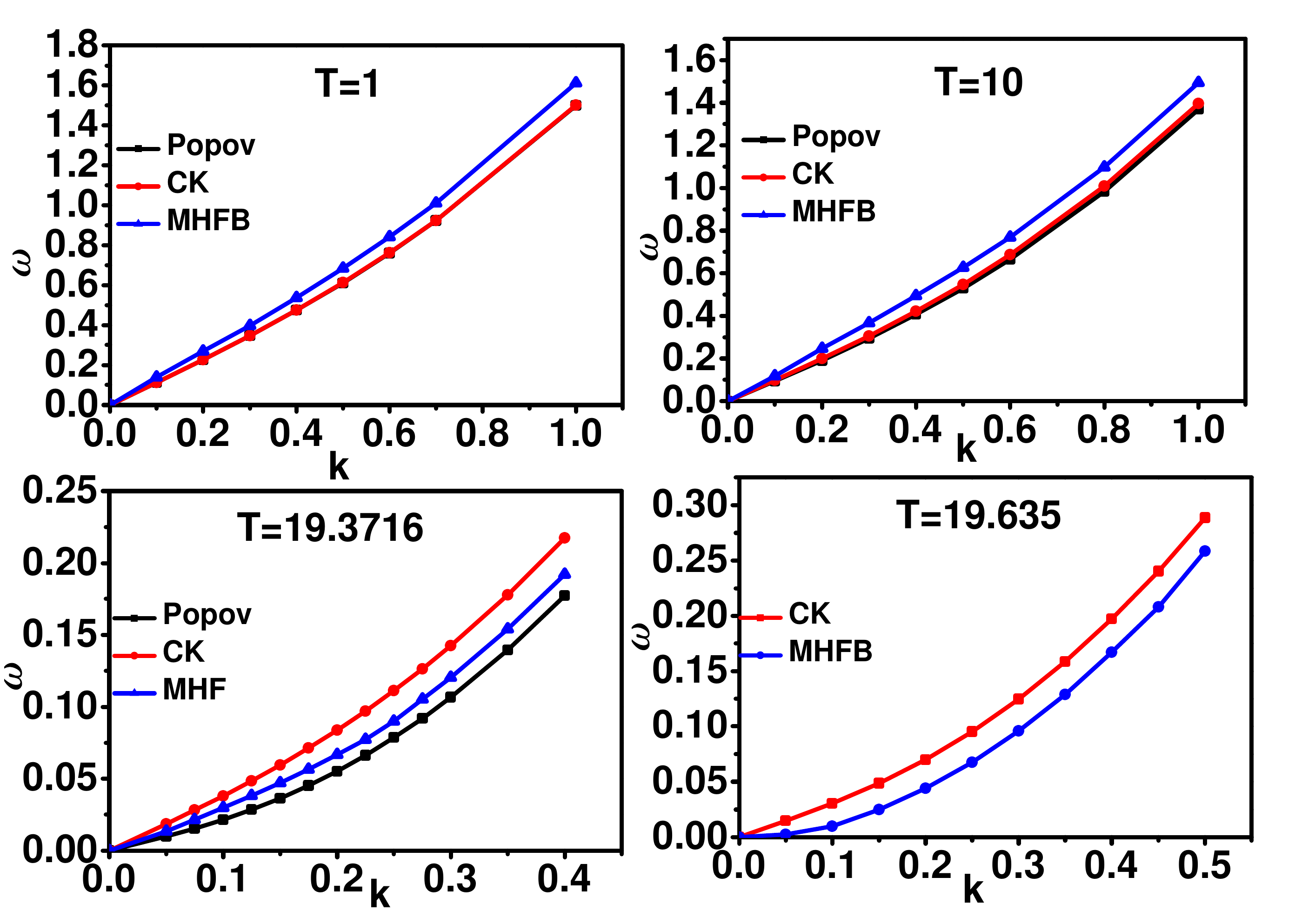}
\caption{The excitation spectrum by Popov theory, Cooper-Kita theory and
Modified HFB when n=5 , a=0.005 for different temperature; $T_0=19.3716,
T_c=19.635$. Kita and Cooper's theories get the same excitation spectrum. At T=1, Cooper and Kita theory is very close to Popov theory.}
\label{fig:es}
\end{figure}

\begin{figure}[tbp]
\centering
\includegraphics[width=0.95\linewidth]{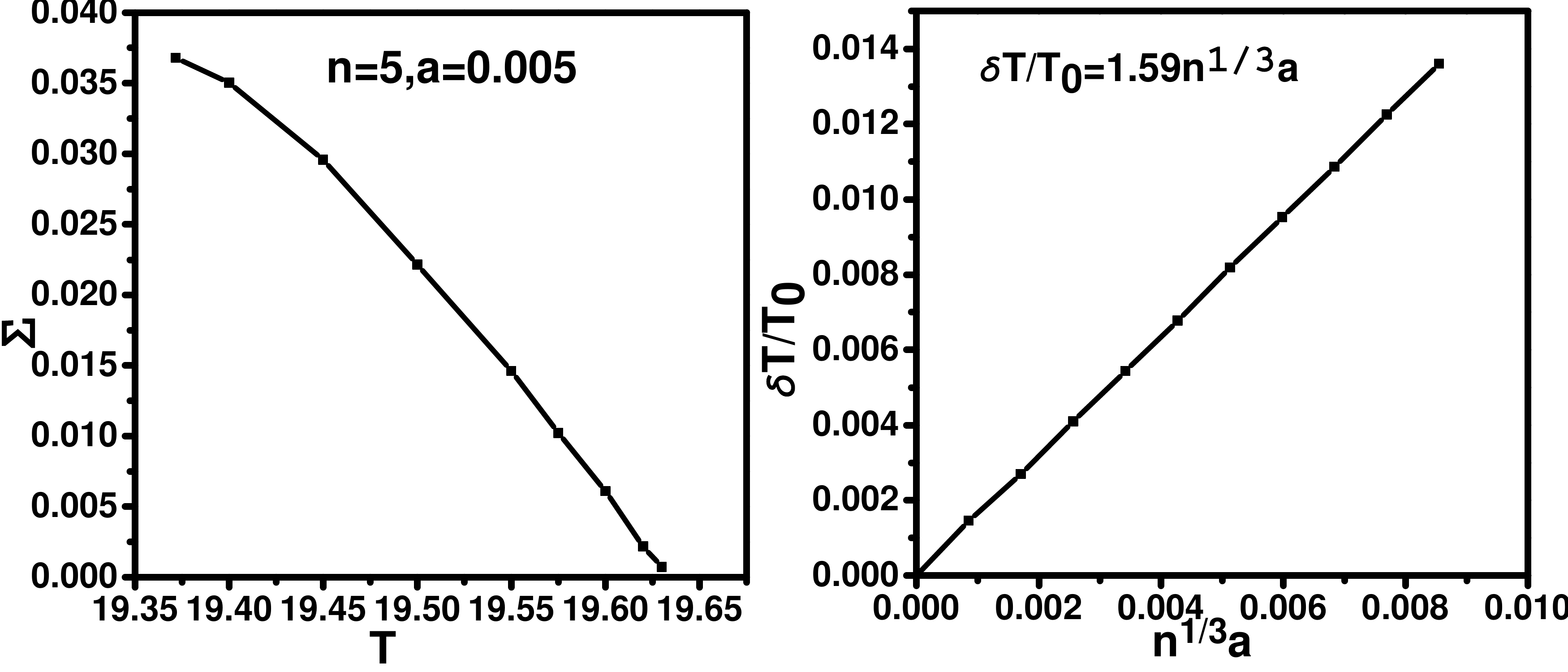}
\caption{(a)the $\Sigma-T$ where $\Sigma$ is fitted by $\protect\omega=%
\protect\sqrt{k^4+2 \Sigma k^2}$; It is clear that the dispersion of energy
spectrum changes to quadratic at $T_c$, n=5, a=0.005. (b)the linear fit of $%
\frac{T_c-T_0}{T_0}=c n^{1/3} a$. We get c=1.59 with R-square=0.99998 }
\label{tc}
\end{figure}

The spectral weight of quasi particle is plotted in FIG.\ref{fig:sw}. The
quasi particle peak is broadened and the quasi particle has finite lift time
caused by the fluctuation effect.

At low temperature, the result of Modified HFB shows discrepancy with Popov
theory and Cooper and Kita theory when interaction is strong, though it coincides with
those theories at weakly interacting limit at low temperature. It has been shown that in one dimension, as interaction
becomes large, the MHFB shows significant difference with Bogoliubov theory and  is
in very good agreement with the exact result at zero temperature\cite{QL}. So we expect that
 systems in three dimension which are less fluctuated than in one dimension,
MHFB will also give quite accurate result. From  FIG.\ref{fig:es}, it is obvious that at low temperature MHFB
gives modifications to Popov theory while Kita and Cooper's theories are close to Popov theory, which can be tested in further experiment
with strong interaction.

$T_{c}$ is the end of the Broken Phase and
actually at this point the linear dispersion of phonon spectrum disappears
and the excitation spectrum becomes $\omega =k^{2}$, which indicates that
this is the critical point of a second order phase transition.By linear fit, we show that $T_{c}$ has a positive shift in comparison with
the idea gas with coefficient $\c=1.59$, while Kita and Cooper theory get $c=2.33$.

In conclusion, we calculate the excitation spectrum of BEC non-perturbatively
with Modified Hartree-Fock-Bogoliubov Theory. Our method is both conserving
and gapless and is valid at the whole temperature regime up to critical
temperature. Our theory predicts a second order phase transition with a
increased critical temperature compared with idea gas $\frac{T_{c}-T_{0}}{%
T_{0}}=1.59n^{1/3}a$. It is different from Popov and Kita and Cooper theory at low
temperature when interaction is strong and it significantly differs from Kita and Cooper
theory near $T_c$. The damping of quasi particle is obtained in our theory while quasi particle
in Popov, Kita and Cooper theory has infinite life time because of the missing of
higher order diagrams. Modified HFB is the simplest Improved $\Phi $-Derivable
Theory. However, the Improved $\Phi $-Derivable Theory can be generalized to higher order.

We thank Professor B. Rosenstein, Professor Lan Yin and Doctor Qiong Li for
valuable discussions. The work is supported by National Natural Science
Foundation (Grant No. 11274018).

\bibliographystyle{apsrev4-1}

\end{document}